\documentclass{elsart4}
\usepackage{graphicx}
\usepackage{dcolumn}
\usepackage{bm}
\usepackage{amsmath}
\usepackage{amsfonts}
\usepackage{amssymb}
\usepackage{srcltx}%
\begin{document}

\begin{frontmatter}

\title{Study of the Kondo and high temperature limits of the slave boson and X-boson methods\thanksref{CNPq}
}
\thanks[CNPq]{We acknowledge the financial support of the CNPq (Brazilian National 
Research Council) and  FAPERJ (Rio de Janeiro State Research Foundation) from the
grant ``Primeiros Projetos''}

\author[IfUFF]{Lizardo H C M Nunes},
\author[IfUFF]{M. S. Figueira},
\ead{figueira@if.uff.br}
\author[IfUNICAMP]{M. E. Foglio}
\address[IfUFF]
{Instituto de F\'{\i}sica, Universidade Federal Fluminense, Av. Litor\^{a}nea s/n, $24210-346$ Niter\'oi-RJ, Brasil.}
\address[IfUNICAMP]
{Instituto de F\'{\i}sica, Universidade Estadual de Campinas, Bar\~ao Geraldo $13083-970$ Campinas-SP, Brasil}

\begin{abstract}
In this paper we study the periodic Anderson model, employing both
the slave-boson and the X-boson approaches in the mean field
approximation. We investigate the breakdown of the slave boson at
intermediate temperatures when the total occupation number of
particles $N_{t}=N_{f}+N_{c}$ is keep constant, where $N_{f}$ and
$N_{c}$ are respectively the occupation numbers of the localized
and conduction electrons, and we show that the high temperature
limit of the slave boson is $N_{f}=N_{c}=N_{t}/2$. We also compare
the results of the two approaches in the Kondo limit and we show
that at low temperatures the X-boson exhibits a phase transition,
from the Kondo heavy Fermion (K-HF) regime to a local moment
magnetic regime (LMM).
\end{abstract}

\end{frontmatter}

\section{Introduction}
\label{int}

The PAM is one of several models of correlated electrons that has been very
useful to describe important physical systems, like transition metals and
heavy fermions (HF). Indeed, heavy fermion materials present a great variety
of ground states: antiferromagnetic (UAgCu$_{4}$, UCu$_{7}$), superconducting
(CeCu$_{2}$Si$_{2}$, UPt$_{3}$), Fermi liquids (CeCu$_{6}$, CeAl$_{3}$) and
Kondo insulators (Ce$_{3}$B$i_{4}$P$t_{3}$, YbB$_{12}$)
\cite{Grewe91,Fulde88,HewsonBook,Aeppli92}. A uniform Curie-like magnetic
susceptibility at high temperature, a common feature of these compounds, is
related to the fact that they contain elements with incomplete $f$-shells,
like $Ce$ and $U$. As the temperature decreases to a certain range, the system
presents a temperature independent uniform susceptibility (Pauli
susceptibility), signaling the quenching of the localized magnetic moments of
the $f$-states, and resembling the behavior of the single-impurity Kondo
problem \cite{MucioBook}. The consistent description of the overall properties
of the heavy fermions is achieved by the competition between the Kondo effect,
dealing with the quenching of the localized magnetic moments, and the
Ruderman, Kittel, Kasuya, Yosida (RKKY) interaction, which favors the
appearance of a magnetic ground state. The basic Hamiltonian that describes
the physics of the HF system should then be a regular lattice of $f$-moments
interacting with an electron gas, and this is the basis of the periodic
Anderson model (PAM), that usually neglects the orbital degeneracy of the
f-electrons, treating them as if they were s-electrons. This interaction
mediates, in higher order, the RKKY magnetic interaction between the f-electrons.

This work compares the results of the mean field slave boson theory (MFSBT)
with those of the mean field X-boson theory (MFXBT) in the Kondo and in the
high temperature limits of the PAM
\cite{Coleman84,Coleman87,Franco02a,Franco02b,Franco03}, taking a constant
number of total electrons per site $N_{t}$. This paper is organized as
follows: in the next section we make a brief revision of the MFSBT and we
calculate $z$, the average number of slave-bosons per site in the system. In
section \ref{X-boson} these results are compared with those obtained by the
X-boson method. In section \ref{KR} we calculate the density of states and the
entropy of the system, and compare the numerical results of the two methods.
In section \ref{Conclusions} we review the main results of the present work.
In  Appendix \ref{Tc} we calculate an analytical expression for the temperature at which
$z\rightarrow0$ in the weak-coupling limit, and in  Appendix \ref{High Temp} we
calculate the MFSBT and MFXBT occupation of $f$-electrons and $c$-electrons in
the high temperature limit.


\section{The slave-boson method}

\label{slave-boson}

In this paper we discuss the periodic Anderson model (PAM) in the limit of
infinite correlation $U\rightarrow\infty$, described by the Hamiltonian
$$
H=\sum_{\mathbf{k},\sigma}E_{\mathbf{k},\sigma}c_{\mathbf{k}%
,\sigma}^{\dagger}c_{\mathbf{k},\sigma}+\sum_{\mathbf{j},\sigma}%
{E}_{f,\mathbf{j,\sigma}}X_{\mathbf{j},\sigma\sigma}+
$$
\begin{equation}
\frac{1}%
{\sqrt{\mathcal{N}_{s}}}\sum_{\mathbf{j},\mathbf{k},\sigma}\left(
v_{\mathbf{k}}e^{i\mathbf{k}.\mathbf{r}_{j}}X_{\mathbf{j},0\sigma}^{\dag
}c_{\mathbf{k},\sigma}+v_{\mathbf{k}}^{\ast}e_{j}^{-i\mathbf{k}.\mathbf{r}%
}c_{\mathbf{k},\sigma}^{\dagger}X_{\mathbf{j},0\sigma}\right)
\,.\label{EqHPAM}%
\end{equation}
The first term is the kinetic energy of the conduction electrons
($c$-electrons), described by the usual Fermi operators, while the second term
is the energy of the localized electrons ($f$-electrons). The last term
represents the hybridization between the $c$-electrons and the $f$-electrons
and we neglect the $f$-$f$ hopping.

An auxiliary boson field is introduced in the slave-boson method
\cite{Coleman84}, and the Hubbard operators at site $\mathbf{j}$ are rewritten
as a product of ordinary boson and fermion operators $X_{\mathbf{j},0\sigma
}^{\dag}\rightarrow f_{\mathbf{j},\sigma}^{\dagger}b_{\mathbf{j}%
}\ ,X_{\mathbf{j},0\sigma}\rightarrow b_{\mathbf{j}}^{\dagger}f_{\mathbf{j}%
,\sigma}$. To eliminate spurious states, only those that preserve the
identity
\begin{equation}
b_{\mathbf{j}}^{\dagger}b_{\mathbf{j}}+\sum_{\sigma}f_{\mathbf{j}\sigma
}^{\dagger}f_{\mathbf{j}\sigma}=I_{\mathbf{j}}, \label{SBconstrain}%
\end{equation}
are considered, and it then follows that $X_{\mathbf{j},00}\rightarrow
b_{\mathbf{j}}^{\dagger}b_{\mathbf{j}}\ $, and $X_{\mathbf{j},\sigma\sigma
}\rightarrow f_{\mathbf{j},\sigma}^{\dagger}f_{\mathbf{j},\sigma}$. At a later
stage Lagrange's method is employed to minimize the thermodynamic potential
with the identity \ref{SBconstrain} as a constrain, and one then has to add
the operator $Q_{\mathbf{j}}\equiv\lambda_{\mathbf{j}}\left(  b_{\mathbf{j}%
}^{\dagger}b_{\mathbf{j}}+\sum_{\sigma}f_{\mathbf{j},\sigma}^{\dagger
}f_{\mathbf{j},\sigma}-I_{\mathbf{j}}\right)  $ to the model Hamiltonian,
where $\lambda_{\mathbf{j}}$ are the Lagrange multipliers. To use the grand
canonical ensemble, we also have to subtract from the Hamiltonian the operator
$\mu\left\{  \sum_{\mathbf{j},\sigma}f_{\mathbf{j},\sigma}^{\dagger
}f_{\mathbf{j},\sigma}+\sum_{\mathbf{k},\sigma}c_{\mathbf{k},\sigma}^{\dagger
}c_{\mathbf{k},\sigma}\right\}  $ where $\mu$ is the chemical potential of the
electrons. The MFSBT is obtained when we neglect the fluctuations of the boson
operators by replacing $b_{\mathbf{j}}$ and $b_{\mathbf{j}}^{\dagger}$\ by
their averages. We assume that the local energies ${E}_{f,\mathbf{j}%
,\sigma}\rightarrow{E}_{f}$ and 
$E_{\mathbf{k},\sigma}\rightarrow E_{\mathbf{k}}$ are site independent and that the
hybridization constant is equal to $V$. Considering translational invariance
we can write $\left\langle b_{\mathbf{j}}^{\dagger}\right\rangle =\left\langle
b_{\mathbf{j}}\right\rangle =\sqrt{z}$, we then obtain the transformed Hamiltonian
$$
H_{\mbox{\scriptsize{S-B}}} =\sum_{\mathbf{k},\sigma}{\epsilon}_{\mathbf{k}%
}\ c_{\mathbf{k},\sigma}^{\dagger}c_{\mathbf{k},\sigma}+\sum_{\mathbf{k}%
,\sigma}\tilde{{E}}_{f}\ f_{\mathbf{k},\sigma}^{\dagger}f_{\mathbf{k},\sigma}+
$$
\begin{equation}
\sum_{\mathbf{k},\sigma}\tilde{V}\ (c_{\mathbf{{k},\sigma}}^{\dagger
}\ f_{\mathbf{k},\sigma}+c_{\mathbf{k},\sigma}\ f_{\mathbf{k},\sigma}%
^{\dagger})+\mathcal{N}_{s}\lambda\left(  z-1\right)  \,, \label{EqSBH}%
\end{equation}
where ${\epsilon}_{\mathbf{k}}={E}_{\mathbf{k}}-\mu$ and the localized
electrons acquire a renormalized energy $\tilde{{E}}_{f}=\epsilon_{f}+\lambda$ with 
$\epsilon_{f}=E_{f}-\mu$, and this Hamiltonian describes a simple uncorrelated Anderson
lattice with renormalized $\tilde{V}\rightarrow\sqrt{z}\ V$ and $\tilde{{E}%
}_{f}$. This model has the following Green's functions (GF) \cite{Newns87}:%
$$
{\ \mathcal{G}}_{\sigma}^{f}(\mathbf{k},\omega) \equiv\ll f_{\mathbf{k}%
,\sigma};f_{\mathbf{k},\sigma}^{\dagger}\gg_{\omega}\ =
$$
\begin{equation}
\frac{-\left(
i\omega-\epsilon_{\mathbf{k}}\right)  }{\left(  i\omega-\tilde{E_{f}}\right)  \left(
i\omega-\epsilon_{\mathbf{k}}\right)  -|\tilde{V}|^{2}}, \label{EqSBGf}%
\end{equation}
with similar expressions for the others Green's functions ${\mathcal{G}}_{\sigma}^{cf}(\mathbf{k}%
,\omega)=\ll c_{\mathbf{k},\sigma};X_{\mathbf{k},0\sigma}^{\dagger}\gg
_{\omega}$ and ${\mathcal{G}}_{\sigma}^{c}(\mathbf{k},\omega)=\ll
c_{\mathbf{k},\sigma};c_{\mathbf{k},0\sigma}^{\dagger}\gg_{\omega},$ all of
them having poles at $\omega_\pm$:%
\begin{equation}
\omega_\pm=\frac{1}{2}\left(  \epsilon_{\mathbf{k}}+\tilde{E_{f}}\pm\sqrt
{(\epsilon_{\mathbf{k}}-\tilde{{E}}_{f})^{2}+4\tilde{V}^{2}}\right)  \,.
\label{EqSBPoles}%
\end{equation}
The occupations per site $N_{f}\equiv\left(  1/\mathcal{N}_{s}\right)
\sum_{\sigma,\mathbf{k}}\langle f_{\mathbf{k,\sigma}}^{\dagger}f_{\mathbf{k}%
\sigma}\rangle$, and $N_{c}\equiv\left(  1/\mathcal{N}_{s}\right)
\sum_{\sigma,\mathbf{k}}\langle c_{\mathbf{k},\sigma}^{\dagger}c_{\mathbf{k}%
,\sigma}\rangle$, as well as the average $N_{cf}\equiv\left(  1/\mathcal{N}%
_{s}\right)  \sum_{\sigma,\mathbf{k}}\langle c_{\mathbf{k},\sigma
}f_{\mathbf{k},\sigma}^{\dagger}\rangle$, can be obtained from these Green's functions.

For a certain region of the parameter space there is a \ \textquotedblleft
condensation temperature\textquotedblright\ $T_{\mbox{\scriptsize{cond}}}$
where $z\rightarrow0$ . This $T_{\mbox{\scriptsize{cond}}}$ was calculated in the Appendix
\ref{Tc} in the weak-coupling limit (cf. (\ref{EqSBTK})), and when we employ
the simple density of states \emph{ }%
\begin{equation}
\rho(E_{\mathbf{k}})=\left\{
\begin{array}
[c]{l}%
(2D)^{-1},\mbox{ for }-D\leq E_{\mathbf{k}}\leq D\\
0\qquad\mbox{\  ,\ \  otherwise }
\end{array}
\right.  \label{EqSquareBand}%
\end{equation}
it can be re-expressed as \emph{ }
\begin{equation}
k_{B}T_{\mbox{\scriptsize{cond}}}\sim\left(  D-\mu\right)  \exp\left\{
-1/(N_{\sigma}\rho_{c}(\mu)J_{K})\right\}  \,. \label{EqSBTKI}%
\end{equation}
In this equation $J_{K}$ is the Kondo coupling obtained via the
Schrieffer-Wolff transformation \cite{Schrieffer66}, $\rho_{c}(\mu)$ is the
density of states of the $c$-electrons at the chemical potential, $D-\mu$ can
be interpreted as a cut-off energy measured from the chemical potential and
$N_{\sigma}$ is the number of spin components per state $\mathbf{k}$. This is
the well-known expression for the single-impurity Kondo temperature
\cite{HewsonBook} and $T_{\mbox{\scriptsize{cond}}}$ is often identified to be
the actual Kondo temperature of the lattice in the mean-field slave-boson
theory. However, as shall be discussed below, such identification is not
appropriate for all \ the values of the total number of electrons $N_{t}%
=N_{f}+N_{c}$. A drawback of the MFSBT is that the formalism present a
discontinuity at $T=T_{\mbox{\scriptsize{cond}}}$
\cite{Coleman87,Franco02a,Burdin00}, which defines a spurious second-order
phase transition that is not observed experimentally. Indeed, at the
condensation temperature we have $N_{f}=1$, so that $z=1-N_{f}=0$ and
$\tilde{V}\equiv\sqrt{z}\ V=0$, decoupling the two bands of the model Hamiltonian.

To analyze this behavior of the MFSBT, we calculate $z$ as a function of the
temperature $T$ for several values of $N_{t}$. In our numerical calculations
we employ $E_{f}=$ $-0.15D$ as the bare localized $f$-energy and $V=0.20D$ as
the hybridization respectively, and use the density of states given by
(\ref{EqSquareBand}).

\begin{figure}[tbh]
\centerline {\ \hspace{-0.4cm}
\includegraphics[width=0.38\textwidth,
angle=-90]
{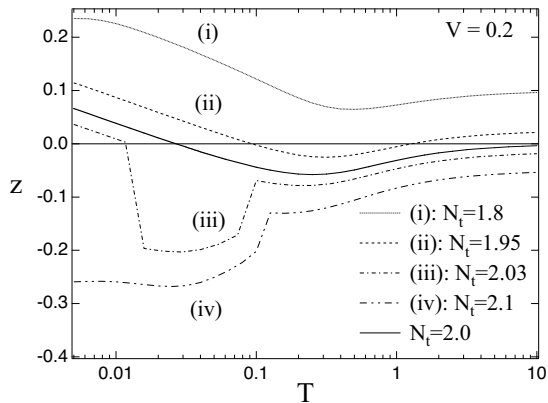} }\caption{The temperature dependence of slave-boson parameter
$z$ for several $N_{t}$. $E_{f}=-0.15D$ and the hybridization parameter $V$ is
$V=0.20D$. All the energies are given in units of $D$. }%
\label{FigSBV02zxT}%
\end{figure}

There is a belief in literature that, for the lattice case, the collapse of the
quasi-particles bands in the MFSBT, occurs  only at very high temperatures, in contrast with the impurity case where this breakdown occurs at the condensation temperature $T_{cond}$ (see reference \cite{HewsonBook} - pg. $343$). But the results  summarized in figure \ref{FigSBV02zxT},
shows  that for the lattice case, the MFSBT present a more complex  behavior. We identify  four distinct regimes for $z$ as a function of $T$. (i) For
$N_{t}=1.8$, the parameter $z$ is positive for all $T$, and the system never
reaches a temperature $T=T_{\mbox{\scriptsize{cond}}}$. (ii) For $N_{t}=1.95$,
$z$ crosses the horizontal axis in two points, defining two condensation
temperatures and the reason for this behavior is that $N_{f}=N_{c}$ in the
high temperature limit (cf. the Appendix \ref{High Temp}), so that $z=1-N_{f}$ becomes
again positive in this limit. (iii) For $N_{t}\gtrsim2$ ($N_{t}=2.03$ in the
figure), $z$ is negative for all $T>T_{\mbox{\scriptsize{cond}}}$ and positive
otherwise. (iv) Finally, in the high occupations regime ($N_{t}=2.1$ in the
figure), $z$ is negative for all temperatures. However, note that the system
never reaches $T_{\mbox{\scriptsize{cond}}}$ for $N_{t}\lesssim1.9$, so that
$T_{\mbox{\scriptsize{cond}}}$ cannot be identified to be the Kondo
temperature of the lattice. The $T=T_{\mbox{\scriptsize{cond}}}$ is defined in
a unique way in the impurity case, because all the calculations are performed
with a constant chemical potential $\mu$, but in the lattice case it is the
total number of particles that should be kept constant, and this modifies the
behavior of the occupation numbers at high temperatures (cf. the Appendix  \ref{High Temp}).

\section{The X-boson method}
\label{X-boson}

In this section we present a brief review of the X-boson method previously
developed \cite{Franco02a,Franco02b,Franco03}. In the PAM Hamiltonian given in
\ref{EqHPAM}, we have employed the Hubbard operators \cite{Hubbard2} to
project out the f-levels with double occupancy from the space of the local
states. Since the Hubbard operators do not satisfy the usual commutation
relations, the diagrammatic methods based on Wick's theorem are not
applicable, and one has to use instead \cite{Figueira94} the product rule:
$X_{f,ab}.X_{f,cd}=\delta_{b,c}X_{f,ad}\,.$ The identity $I_{\mathbf{j}}$ in
the restrained space of local states at site $\mathbf{j}$ is then
$X_{\mathbf{j},00} + X_{\mathbf{j},\sigma\sigma} +X_{\mathbf{j},\overline
{\sigma}\overline{\sigma}} = I_{\mathbf{j}} $, where $\overline{\sigma} =
-\sigma$, and we shall call \textquotedblleft completeness\textquotedblright%
\ the average
\begin{equation}
\left\langle X_{\mathbf{j},00} \right\rangle + \left\langle X_{\mathbf{j},
\sigma\sigma} \right\rangle + \left\langle X_{ \mathbf{j},\overline{\sigma
}\overline{\sigma} } \right\rangle =1 . \label{EqCompleteness}%
\end{equation}
We shall now consider the cumulant expansion \cite{Figueira94} of the relevant
GFs. We use the \textquotedblleft chain approximation\textquotedblright%
\ (CHA), that contains all the possible diagrams with only second order
cumulants and it is still fairly simple to handle. Although the exact GF
satisfy completeness (i.e. (\ref{EqCompleteness})), the different approximate
GFs do not usually have this property, and this is the case with the CHA
\cite{Figueira96}. We introduce the average value $R=\left\langle
X_{\mathbf{j},00}\right\rangle $, that by the translational invariance is
independent of the site $\mathbf{j}$, and is analogous to the mean-value $z$
employed in the slave-boson technique. Following the MFSBT we use $R$ as a
variational parameter, and satisfy completeness by minimizing the
thermodynamic potential with (\ref{EqCompleteness}) as a constraint. We use
again Lagrange's method, and to enforce the \textquotedblleft
completeness\textquotedblright\ in this mean-field approximation we have to
add the operator $Q\equiv\Lambda\left[  \sum_{\mathbf{k}}\left(
R+\sum_{\sigma}X_{\mathbf{k},\sigma\sigma}-1\right)  \right]  $ to the
Hamiltonian employed to calculate the thermodynamic potential. For the grand
canonical ensemble we then have to use the following transformed Hamiltonian%
$$
H_{\scriptsize{X-B}}=\sum_{\mathbf{k},\sigma}\epsilon_{\mathbf{k}%
}c_{\mathbf{k},\sigma}^{\dagger}c_{\mathbf{k},\sigma}+\sum_{\mathbf{k},\sigma
}\tilde{E}_{f}X_{\mathbf{k},\sigma\sigma}+
$$
\begin{equation}
V\sum_{\mathbf{{k}},\sigma}\left(  X_{\mathbf{k},0\sigma}^{\dagger
}c_{\mathbf{{k}},\sigma}+c_{\mathbf{{k}},\sigma}^{\dagger}\ X_{\mathbf{k}%
,0\sigma}\right)  +\mathcal{N}_{s}\mathcal{\ }\Lambda\ \left(  R-1\right)  \,,
\label{EqXBHa}%
\end{equation}
with the renormalized energies $\epsilon_{\mathbf{k}}=E_{k}-\mu$, $\tilde{{E}}_{f}=\epsilon_{f}+\Lambda$ with 
$\epsilon_{f}=E_{f}-\mu$ for the $c$ and $f$-electrons respectively, as obtained in the MFSBT. With this Hamiltonian the CHA
gives
$$
{\mathcal{G}}_{\sigma}^{f}(\mathbf{k},\omega)\equiv\ll X_{\mathbf{k}%
,0\sigma};X_{\mathbf{k},0\sigma}^{\dagger}\gg_{\omega}=
$$
\begin{equation}
\frac{-D_{\sigma
}\ \left(  i\omega-\epsilon_{\mathbf{k}}\right)  }{\left(  i\omega-\tilde{E_{f}%
}\right)  \left(  i\omega-\epsilon_{\mathbf{k}}\right)  -|V_{\sigma}(\mathbf{k}%
)|^{2}D_{\sigma}}, \label{EqXBGf}%
\end{equation}
where $D_{\sigma}\equiv\langle X_{\mathbf{j},00}+X_{\mathbf{j},\sigma\sigma
}\rangle$, and similar expressions for ${\mathcal{G}}_{\sigma}^{cf}%
(\mathbf{k},\omega)=\ll c_{\mathbf{k},\sigma};X_{\mathbf{k},0\sigma}^{\dagger
}\gg_{\omega}$ and ${\mathcal{G}}_{\sigma}^{c}(\mathbf{k},\omega)=\ll
c_{\mathbf{k},\sigma};c_{\mathbf{k},0\sigma}^{\dagger}\gg_{\omega}$
\cite{Franco02a}. The poles of the GFs are now given by
\begin{equation}
\omega_\pm^{\sigma}=\frac{1}{2}\left(\epsilon_{\mathbf{k}}+\tilde{E_{f}}\pm
\sqrt{(\epsilon_{\mathbf{k}}-\tilde{{E}}_{f})^{2}+4D_{\sigma}V^{2}}\right)  \,,
\label{EqXBPoles}%
\end{equation}
which differ formally from (\ref{EqSBPoles}) only by the presence of the factor
$D_{\sigma}$.

Since we are constrained to the Hilbert subspace where the \textquotedblleft
completeness\textquotedblright\ relation (\ref{EqCompleteness}) is satisfied,
we find in the paramagnetic case ( $\sum_{\mathbf{k}}\langle X_{\mathbf{k}%
,\sigma\sigma}\rangle=\sum_{\mathbf{k}}\langle X_{\mathbf{k},\overline{\sigma
}\overline{\sigma}}\rangle$) that $D_{\sigma}=1-(N_{f}/2)$. The minimization
of the thermodynamic potential with respect to $R$ gives the Lagrange
multiplier $\Lambda$ \cite{Franco02a,Franco02b}:
\begin{equation}
\Lambda=-V^{2}\frac{1}{\mathcal{N}_{s}}\sum_{\mathbf{k},\sigma}\frac
{n_{F}\left(  \omega_{+}^{\sigma}\right)  -n_{F}\left(  \omega_{-}^{\sigma
}\right)  }{\omega_{+}^{\sigma}-\omega_{-}^{\sigma}}\,, \label{EqXBLambda}%
\end{equation}
which is analogous to the slave-boson result. Note that although the GFs of
the MFXBT are very similar to the uncorrelated ones ($U=0$), they cannot be
reduced to them by any change of scale, except for $D_{\sigma}\rightarrow1$,
when we recover the slave-boson GFs for $V\rightarrow\sqrt{z}V$. Indeed, the
strong correlations effects present in the system appear naturally (in a mean
field way) in the MFXBT through the quantity $D_{\sigma}$, which enforces the
condition $N_{f}\leq1$ and $R>0$ for all temperatures and occupations. The
main advantage of the present treatment is that eliminates the spurious phase
transition appearing in the MFSBT when $z=0$, as well as the regions with
$z<0$.

\begin{figure}[tbh]
\centerline {\
\includegraphics[width=0.33\textwidth,
angle=90]
{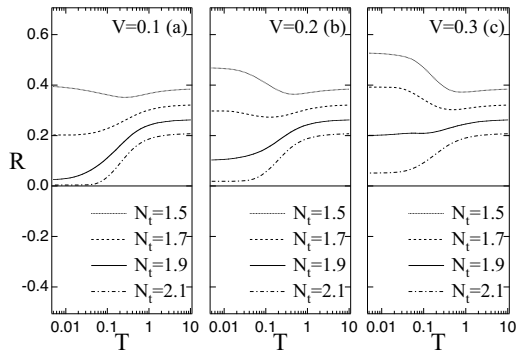} }\caption{The temperature dependence of $R$ for several $N_{t}$.
Other parameters are the same as in Fig. \ref{FigSBV02zxT}.}%
\label{FigXBRxT}%
\end{figure}
In analogy to the discussion performed in the previous section we calculate
the quantity $R$ (equivalent to the MFSBT parameter $z$)$\ $as a function of
$T$ for several $N_{t}$ within the X-boson approach, and with the density of
states given at (\ref{EqSquareBand}). We plot our results for $V=0.1$,
$V=0.2$, and $V=0.3$ in figure \ref{FigXBRxT}, and they show the same tendency
displayed in figure \ref{FigSBV02zxT} by the MFSBT: for a given $N_{t}$ there
is an enhancement of $R$ as $V$ increases in the low-temperature regime.
Differently from the MFSBT, $R$ is positive for every range of temperatures
and occupations, since $N_{f}\leq1$, as follows from the analysis in the Appendix
\ref{High Temp}.

In the present paper we adopt a schematic classification proposed by Varma
\cite{Varma85} and recently reintroduced by Steglich et.al.
\cite{Grewe91,Steglich94}, which illustrates the competition between magnetic
order and Fermi liquid formation. This classification is given in terms of the
dimensionless coupling constant for the exchange between the local $f$ spin
and the conduction-electron spins, given by $g=\rho_{c}(\mu)|J_{K}|$. The
$J_{K}$ is the Kondo coupling constant, connected to the parameters of the PAM
via the Schrieffer-Wolff transformation \cite{Schrieffer66} that gives
$J_{K}=2V^{2}/|E_{f}-\mu|$ when $U\rightarrow\infty$. Within the MFSBT or the
MFXBT we then have that
\begin{equation}
g=\rho_{c}(\mu)|J_{K}|\ \sim\frac{V^{2}}{D|\epsilon_{f}|}\, , \label{Eqg}%
\end{equation}
where for simplicity we take $\rho_{c}(\mu)=1/(2D)$. The qualitative behavior of
the exemplary Ce-based compounds is related to this parameter as follows: when
$g>1$, the compound presents an intermediate valence (IV) behavior, while for
$g<1$ it is in a heavy fermion Kondo regime (HF-K). There exists a critical
value $g_{c}$ at which the Kondo and the RKKY interactions have the same
strength, and non Fermi-liquid (NFL) effects have been postulated when
$g=g_{c}$. For $g_{c}<g<1$, the magnetic local moments are not apparent at
very low temperatures and the system presents a Fermi liquid behavior, while
for $g<g_{c}$ the system is in the local magnetic moment regime (LMM). We
point out that the parameter $g$ classifies the regimes of the PAM only in a
very qualitative way.

\section{The Kondo temperature}
\label{TK}

In this section we calculate the Kondo temperature ($T_{K}$) of the PAM
following a scheme proposed by Bernhard and Lacroix \cite{Bernhard99}, which
defines $T_{K}$ as the minimum of the temperature derivative of $N_{cf}$,
where the average $N_{cf}$ is obtained from ${\ \mathcal{G}}_{\sigma}^{cf}$
and measures the \textquotedblleft transference\textquotedblright\ of
electrons from the localized levels to the conduction band and vice versa.
Note that $T_{K}$ is not a true order parameter, but in fact establishes a
crossover temperature between the two regimes of the Anderson lattice in the
nonmagnetic case: a low temperature regime with no local magnetic moments,
also referred as the Kondo regime of the system, and a high-temperature regime
characterized by the presence of disordered local magnetic moments.

\begin{figure}[tbh]
\centerline {
\includegraphics[width=0.33\textwidth,
angle=90]
{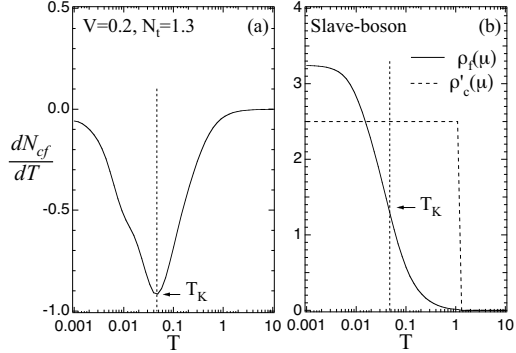}  } \caption{The temperature dependence of (a)
$dN_{cf}/dT$, (b) $\rho_{f}(\mu)$, and ${\rho^{\prime}}_{c}(\mu)=5\rho_{c}%
(\mu)$ in the MFSBT for $V=0.2$ and $N_{t}=1.3$. The vertical dashed line
indicates $T_{K}$. Other parameters are the same as in Fig. \ref{FigSBV02zxT}%
.}%
\label{FigSBV02NT13DTGNFCandRF0xT}%
\end{figure}
As we have shown in Section \ref{slave-boson}, $T_{\mbox{\scriptsize{cond}}}$
is never reached within the MFSBT in the low occupations regime, and therefore
the Kondo temperature of the system cannot be identified with
$T_{\mbox{\scriptsize{cond}}}$. However, for every hybridization parameter
investigated, our data always presents a global minimum of $dN_{cf}/dT$ in the
low occupation regime, which defines $T_{K}$.
The temperature dependence of
$dN_{cf}/dT$ for $V=0.2$ and $N_{t}=1.3$ is shown in Fig.
\ref{FigSBV02NT13DTGNFCandRF0xT}.a, where the vertical dashed line indicates
$T_{K}$. Within the MFSBT, the parameter $g$ is $0.645$ at $T=0.001$ for this
occupation, and this value corresponds to the HF regime. This is corroborated
by the temperature dependence of $\rho_{f}(\mu)$ and $\rho_{c}(\mu)$ presented
in Fig. \ref{FigSBV02NT13DTGNFCandRF0xT}.b, since $\mu$ lies in the vicinity
of the Kondo resonance.
\begin{figure}[tbhtbh]
\centerline{
\includegraphics[width=0.33\textwidth,
angle=90]
{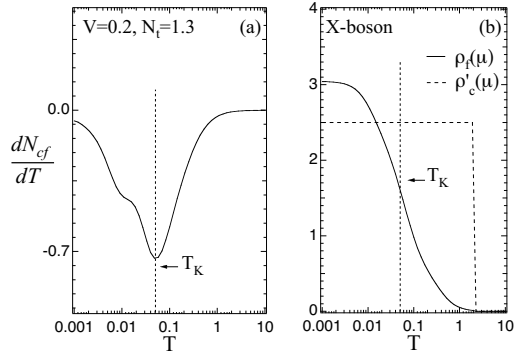}  } \caption{The temperature dependence of (a)
$dN_{cf}/dT$, (b) $\rho_{f}(\mu)$, and ${\rho^{\prime}}_{c}(\mu)=5\rho_{c}%
(\mu)$ in the MFXBT for $V=0.2$ and $N_{t}=1.3$. The vertical dashed line
indicates $T_{K}$. Other parameters are the same as in Fig. \ref{FigSBV02zxT}%
.}%
\label{FigXBV02NT13DTGNFCandRF0xT}%
\end{figure}
A similar analysis applies to the results displayed in Fig.
\ref{FigXBV02NT13DTGNFCandRF0xT} for the MFXBT, where $g=0.607125$ at
$T=0.001$. These results indicate that both the X-boson and the slave-boson
approaches provide almost the same quantitative description for the system in
the low occupation regime.

\begin{figure}[tbh]
\centerline { \includegraphics[width=0.33\textwidth,
angle=90]
{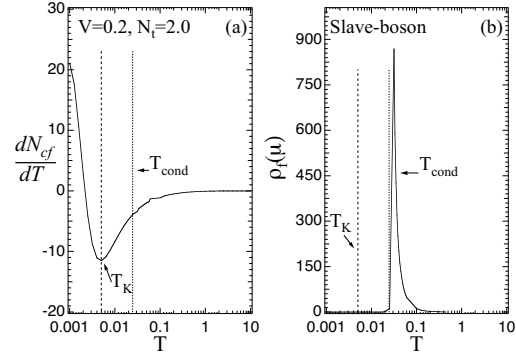} } \caption{The temperature dependence of (a)
$dN_{cf}/dT$ and (b) $\rho_{f}(\mu)$ in the MFSBT for $V=0.2$ and $N_{t}=2.0$.
The vertical dashed line indicates $T_{K}$ and the vertical short dashed line
indicates $T_{\mbox{\scriptsize{cond}}}$. Other parameters are the same as in
Fig. \ref{FigSBV02zxT}.}%
\label{FigSBV02NT20DTGNFCandRF0xT}%
\end{figure}
For occupations in which $T=T_{\mbox{\scriptsize{cond}}}$ exists
in the slave-boson approach, our data has always a global minimum
for $dN_{cf}/dT$ in the interval $T<T_{\mbox{\scriptsize{cond}}}$,
as can be seen in Fig. \ref{FigSBV02NT20DTGNFCandRF0xT}.a for
$N_{t}=2.0$ and $V=0.2$. At $T=0.001$, we have that $g=46.17$ in
the slave-boson approach, corresponding to an IV regime. Indeed,
as can be seen in Fig. \ref{FigSBV02NT20DTGNFCandRF0xT}.b,
the system is an insulator and
several Kondo insulators present the same type of behavior.
\begin{figure}[tbhtbh]
\centerline{
\includegraphics[width=0.33\textwidth,
angle=90]
{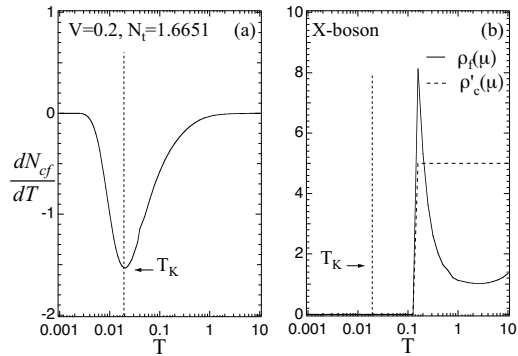}  } \caption{The temperature dependence of
(a) $dN_{cf}/dT$, (b) $\rho_{f}(\mu)$, and ${\rho^{\prime}}_{c}(\mu
)=10\rho_{c}(\mu)$ in the MFXBT for $V=0.2$ and $N_{t}=1.6651$. The vertical
dashed lines indicates $T_{K}$. Other parameters are the same as in Fig.
\ref{FigSBV02zxT}.}%
\label{FigXBV02NT166DTGNFCandRF0xT}%
\end{figure}
These data should be compared with the results obtained for $N_{t}=1.6651$ and
$V=0.2$ in the MFXBT, where the system is also an insulator at this
occupation, in which $g=2.14$ at $T=0.001$. From Fig.
\ref{FigXBV02NT166DTGNFCandRF0xT} we can then infer that both the X-boson and
the slave-boson approaches yield the same qualitative description for the
system, but in the X-boson approach we do not have the spurious second-order
phase transition presented by the MFSBT at $T=T_{\mbox{\scriptsize{cond}}}$.

\section{Kondo Regime}
\label{KR}

To investigate the Kondo regime of the system, we vary the bare
energy $E_f$ from the empty dot regime ($E_f >\mu $) to the
extreme Kondo limit, where $E_f <<\mu $. As $E_f$ goes to the
Kondo limit the charge fluctuations are suppressed, while the spin
fluctuations become dominant. In the numerical calculations of
this section we always employ the total occupation number $N_{ t }
=1.3 $ and the hybridization $V=0.20$, and we calculate the
chemical potential self-consistently. In the curves of density of
states we give the frequencies with respect to the chemical
potential, so that always is $\mu =0$. Also note that for the
MFSBT we shall plot the ``real" particle density of states,
$\rho^{ \mbox{\scriptsize{real}} }_{ f } ( \mu) = z \rho_{ f } (
\mu)$, where $\rho_{ f } ( \mu) $ is the quasi-particle density of
states described by the usual fermionic operators in the MFSBT.
For simplicity we shall suppress the ``real" superscript along
this section.

\begin{figure}[tbh]
\centerline{
\includegraphics[width=0.36\textwidth,angle=-90]
{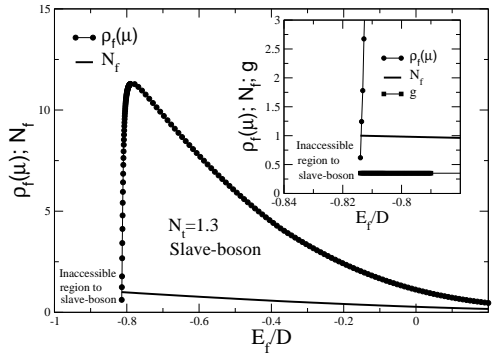} }\caption{The
spectral density $\rho_{ f }( \mu) $ of local states and the total
number of particles $N_{ f } $ as a function of the position of
the bare local level $E_{ f } $ in the MFSBT, for $V=0.2 $ and
$N_{ t }=1.3 $ and $T=10^{-5}$. The inset shows the point where
$\rho_{ f }( \mu) \rightarrow0 $ and $N_{ f }
\rightarrow 1 $. }%
\label{FigSBV02NT13rhofxEf}%
\end{figure}

The MFSBT results at $T=10^{-5}$ are shown in figure
\ref{FigSBV02NT13rhofxEf}. As can be seen in the figure, $\rho_{ f
} (\mu) $ reaches a peak as $E_{ f } $ decreases, denoting the
enhancement of the effective mass of the system and its heavy
fermion behavior due to the Kondo effect. As $E_{ f } $ decreases
even more, the $f $-density of states at the Fermi level falls
steeply. The inset indeed shows that $\rho_{ f } (\mu)
\rightarrow0 $ when the $f $-band occupation $N_{ f } $ approaches
unity, i.e. the Kondo limit in the slave-boson approach, as
already discussed above. Moreover, as $E_{ f } $ decreases even
more, we reach an interval where $z < 0 $, corresponding to
unphysical results  of the MFSBT, i.e. to an inaccessible region
for the slave-boson method.

\begin{figure}[tbh]
\centerline{
\includegraphics[width=0.38\textwidth, angle=-90]
{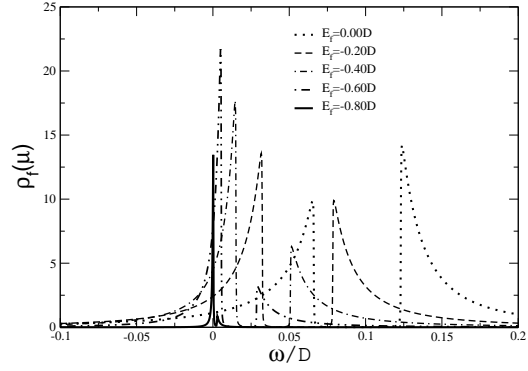}
}\caption{The spectral localized density of states $\rho_{ f }( \omega) $ as a
function of the frequency $\omega$ in the MFSBT employing the same values used
in Fig. \ref{FigSBV02NT13rhofxEf}.}%
\label{FigSBV02NT13rhofxw}%
\end{figure}

Fig. \ref{FigSBV02NT13rhofxw} shows the $f $-band densities of states as a
function of $\omega$ (taking the chemical potential at the origin). The hybridization gap goes to zero at the Kondo limit, and the first peak of the density of states mimics the lattice Kondo peak.

\begin{figure}[tbh]
\centerline{
\includegraphics[width=0.36\textwidth, angle=-90]
{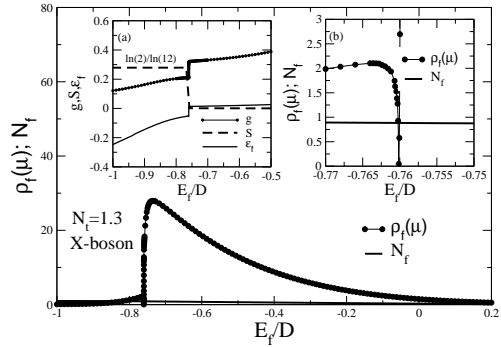}}\caption{The spectral
localized density of states $\rho_{ f }( \mu) $, and $N_{ f } $ as a function
of the position of the localized level $E_{ f } $ in the MFXBT employing the
same values used in figure \ref{FigSBV02NT13rhofxEf}. The inset $(a)$ shows a
detail of the parameter $g$, the entropy $S$ and the localized level $\epsilon_{f}=E_{f}-\mu$ in the region of the
transition from the HF-K to LMM regime. The inset $(b)$ shows the behavior of
the $\rho_{ f }( \mu) $ and $n_{f}$ around the transition.}%
\label{FigXBV02NT13rhofxEf}%
\end{figure}

On the same token, the MFXBT results for $N_{ t } = 1.3 $ and $V = 0.2 $ are
shown in figure (\ref{FigXBV02NT13rhofxEf}). The X-boson results are similar to those obtained in the MFSBT: $\rho_{ f } (\mu)$ reaches a peak as $E_{ f } $ decreases, denoting the effective mass enhancement of the heavy fermions, and if further decreases when the $E_{ f } $ decreases even more.
{\bf In the inset (a) we plot the parameter $g$, the entropy $S$  and the localized level $\epsilon_{f}=E_{f}-\mu$ as function of $E_{f}$, and all these quantities present a steeply but
continuous behavior at $E_{ f } \approx-0.76 $. We identify this
``jump" as the transition from the heavy Fermion Kondo regime (HF-K) to
the local magnetic moment regime (LMM). As pointed out by Steglich \cite{Steglich94}, in real systems, the majority of HF-K compounds is found in the region where $g < g_{c}$ and these systems suffer a magnetic phase transition before both the heavy masses and coherence among the quasiparticles can fully develop.
The results obtained by the X-boson are consistent with this scenario,
but here we do not calculate magnetic solutions of the problem; 
we can only say that in the LMM phase the system presents an effective mass 
which is lower than in the HF-Kondo regime,
as indicated by the specific heat $\gamma$ coefficient calculation \cite{Franco03}.
Considering the Kondo and RKKY energies,
since $k_{B}T_{RKKY} \sim g^{2}$ and $k_{B}T^{cond} \sim \exp(-1/g)$ respectively,
we obtain that the $E_f$ value 
when the Kondo and the RKKY interactions have the same strength, 
occurs at around the $E_{f}$ critical transition value,
what leads us to consider this region as ``associated", in a mean field way, 
to the critical parameter $g_{c}$,
as discussed in the end of the Section \ref{X-boson}. 
Nevertheless, we should be cautious about this result, 
since the X-boson is only a mean field theory and within this formalism
it is not possible to capture all the relevant physics associated to this transition.
It is believed that this transition defines a Quantum Critical Point (QCP) \cite{MucioBook}.
However, given that the X-boson self-energy does not depend on the wave vector,
we cannot take into consideration the RKKY interaction
and we cannot discuss the non-Fermi liquid behavior,
nor find the correct $g_{c}$ value associated with the QCP.
As can be seen in the inset (a), the entropy $S$ per site in the HF-K regime is close to zero, 
signaling the Kondo singlet ground-state;
however, as $\mu$ crosses the hybridization gap and $E_{ f } $
decreases even more, $S $ presents a continuous transition at $E_{ f } \approx-0.76 $.
In the LMM region, $S \rightarrow k_{B} ln(2)/ln(12) $, with $k_{B}=1$ in all the calculations, pointing to the transformation of the singlet of the HF-K regime into a ground state
consisting of a doublet at each site, that could be attributed to a spin $1/2
$ at each site, which is the LMM regime presented by the PAM when $N_{ f }
\rightarrow1 $. This regime cannot be obtained in the slave-boson approach. This transition always appears in the Kondo region, when the chemical potential $\mu$ changes  signal forcing the localized level 
$\epsilon_{f}=E_{f}-\mu$ to enter the LMM region,
as represented in the inset (a) of the Fig. (\ref{FigXBV02NT13rhofxEf}).
In the inset (b)}, we present a detail of $\rho_{ f }( \mu) $, and $N_{ f } $
as a function of the position of the localized level$E_{ f } $. The two
methods give qualitatively different results in this region: while the slave boson
reproduces the impurity Kondo limit the X-boson gives a transition to the LMM
regime, which is essentially a lattice behavior as can be inferred by several
experimental results \cite{Grewe91,Steglich94}.

\begin{figure}[tbh]
\centerline{
\includegraphics[width=0.38\textwidth, angle=-90]
{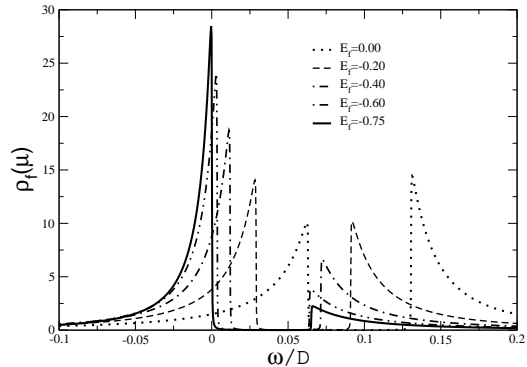}}\caption{The spectral localized
density of states $\rho_{ f }( \omega) $ as a function of the frequency
$\omega$ in the MFXBT employing the same values used in Fig.
\ref{FigSBV02NT13rhofxEf}.}%
\label{FigXBV02NT13rhofxEf_1}%
\end{figure}

Fig. \ref{FigXBV02NT13rhofxEf_1} shows the $f $-band densities of states as a
function of $\omega$ (taking the chemical potential at the origin) in
the MFXBT. As in the slave-boson approach, the density of states exhibits a
double-peak structure for every value of $E_{ f } $, but now the hybridization
gap does not goes to zero as in the slave boson case. The first peak is enhanced at $\mu= 0 $, which mimics again the Kondo peak, and at the same time the second peak loses importance. 

\begin{figure}[tbh]
\centerline{
\includegraphics[width=0.38\textwidth, angle=-90]
{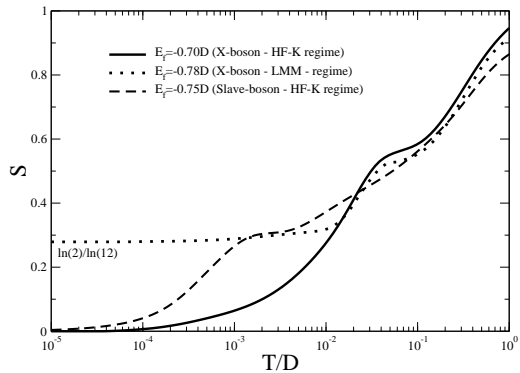}
}\caption{Entropy $S $ as a function of the temperature $T $ in the MFXBT and
MFSBT for several values of $E_{ f } $ using $N_{ t } = 1.3 $ and $V = 0.2 $.}%
\label{FigV02NT13SxT}%
\end{figure}

in Fig. \ref{FigV02NT13SxT} we present the calculation of the entropy
\cite{Franco03} as a function of the temperature for both the slave-boson and the
X-boson methods. For a system with a chemical potential that remains finite
when $T \rightarrow\infty$, the entropy per site would tend to $k_{ B }
\ ln(12) $, corresponding to the dimensionality of the purely local Hilbert
space. The results for $E_{ f } = -0.70 $ in the X-boson and for $-0.75 $  in the
slave-boson approach, show that $S $ tends to zero as $T
\rightarrow 0$, indicating that the system in the ground-state goes to the Kondo singlet. However, in the MFXBT, the entropy $S $ per site for $E_{ f } = -0.78 $ tends in the low temperatures limit to $k_{B}ln(2)/ln(12)$,
indicating a ground state consisting of a doublet at each site.


\section{Conclusions}

\label{Conclusions}

In this paper we have employed both the MFSBT and the MFXBT to study the
paramagnetic case of the PAM in the limit $U=\infty$, and we have compared
their results and predictions.

As we pointed earlier there is a belief in the literature that, for the lattice case, the collapse of the quasi-particles bands in the MFSBT, occurs  only at very high temperatures, in contrast with the impurity case where this breakdown occurs at the condensation  temperature $T_{cond}$ (see reference \cite{HewsonBook} - pg. $343$). But the results  summarized in figure \ref{FigSBV02zxT},
shows  that for the lattice case, the MFSBT present a more complex  behavior.
Within the MFSBT, we have defined a temperature $T_{\mbox{\scriptsize{cond}}}$
at which the average occupation $z$ of the vacuum state is $z=0$, and we have
calculated numerically the temperature dependence of $z$ for a constant total
number of electrons $N_{t}$. Our results show that:
in the low occupations regime, the system never reaches
$T=T_{\mbox{\scriptsize{cond}}}$ at constant $N_{t}$, 
although previous calculations
\cite{Franco02a} for a constant $\mu$ always present the spurious phase
transition at $z=0$;
for a range of occupations $N_{t}\lesssim2$, 
there are two values of $T$ where $ z=0 $: $z$ is positive up to $T=T_{\mbox{\scriptsize{cond}}}$, 
and it then becomes negative for a bounded temperature interval.
As $T$ increases even more, $z=1-N_{f}$ becomes positive again, to satisfy the
high-temperatures result $N_{c}=N_{f}$, shown in Appendix \ref{High Temp};
in the high occupations regime, $z$ is negative for every range of
temperatures.
Finally, from the Appendix \ref{High Temp} we prove that the
high temperature limit of the MFSBT determines completely the strange behavior
of the condensation temperature in the lattice case.

In order to avoid these problems, we study the PAM employing the MFXBT, which
does not have these difficulties of the slave-boson method, giving $N_{f}<1$
for every range of temperatures and occupations, as should be expected for the
PAM in the $U=\infty$ limit. Moreover, the $c$-band and $f$-band densities of
states at the chemical potential are always positive, showing that the X-boson
approach provides physical results for every range of temperatures and
occupations. In the Kondo limit both methods present similar results but the
entropy data in the MFXBT results shows a steeply but continuous transition where the
singlet of the heavy fermion Kondo regime transforms into a ground state
consisting of a doublet at each site, that could be attributed to a spin $1/2
$ at each site, which is the local magnetic moment regime presented by the
PAM. This regime cannot be obtained by the slave-boson approach.

As the final conclusions we can say that the X-boson plays a complementary
role when compared with the slave boson approach which was designed to
describe the Kondo limit of the PAM, but fails as the temperature increases.
On the other hand, since the computational costs of the X-boson is equivalent
to the slave-boson and produces physical results for the PAM at any
temperature or chemical potential, this approach seems useful as a starting
point to study temperature dependent problems of heavy fermion systems like
heavy fermion superconductivity \cite{Nunes03,Nunes03a,Nunes03b}, intermediate
valence systems like the Kondo insulators \cite{Tatiana2000} or the transition
to HF-K to LMM regime \cite{Grewe91,Steglich94}.

\ack We would like to express our gratitude to Prof. M. A. Continentino for
helpful discussions and the financial support of Funda\c c\~ao de Amparo a
Pesquisa do Estado do Rio de Janeiro (FAPERJ) from the grant ``Primeiros
Projetos'' and Conselho Nacional de Desenvolvimento Cient\'{\i}fico (CNPq).

\appendix

\section{The condensation temperature}

\textbf{\label{Tc} } We shall call $T_{\mbox{\scriptsize{cond}}}$ the
temperature at which the system takes the value $z=0$ in the MFSBT. The constrain \ref{SBconstrain} to the Hilbert space gives, $z+{\mathcal{N}_{s}}^{-1}%
\sum_{\mathbf{k},\sigma}\langle f_{\mathbf{k,\sigma}}^{\dagger}f_{\mathbf{k}%
\sigma}\rangle=1$, the $T_{\mbox{\scriptsize{cond}}}$ can be
reached if, and only if, $N_{f}=1$. When $z\rightarrow0$ it
follows $\tilde{V}=\sqrt {z}\ V\rightarrow0$, and from
(\ref{EqSBGf}) we have ${\mathcal{G}}_{\sigma
}^{f}(\mathbf{k},\omega)\equiv-1/\left(
i\omega-\tilde{E_{f}}\right)  \,$, so that
$N_{f}=N_{\sigma}n_{F}(\tilde{E_{f}})$, where $N_{\sigma}=2$ is
the
number of spin components. The relation $N_{f}=1$ then implies $\tilde{E_{f}%
}=0$, and we obtain $T_{\mbox{\scriptsize{cond}}}$ by taking simultaneously
$z=0$ and $\tilde{E_{f}}=0$.

With the same procedure employed in \cite{Franco02a} we obtain%
\begin{equation}
\lambda=-V^{2}N_{\sigma}\int d\epsilon\rho(\epsilon)\,\frac{n_{F}\left(
\omega_{+}\right)  -n_{F}\left(  \omega_{-}\right)  }{\omega_{+}-\omega_{-}%
}\,,\label{EqSBLambda}%
\end{equation}
 where $n_{F}(x)\equiv\left(  1+e^{\beta x}\right)  ^{-1}$ is the Fermi
function. This equation is analogous to the (\ref{EqXBLambda}) in the X-boson
approach. At $T=T_{\mbox{\scriptsize{cond}}}$, we get that $\lambda=\mu-E_{f}%
$, and (\ref{EqSBLambda}) can be rewritten as\textbf{
\begin{equation}
1=\frac{N_{\sigma}V^{2}}{2\left(  \mu-E_{f}\right)  }\int d\epsilon
\,\rho(\epsilon)\frac{1-2n_{F}(\epsilon-\mu)}{\epsilon-\mu}%
\,,\label{EqSBTcond}%
\end{equation}
}which is a self-consistent equation for $T_{\mbox{\scriptsize{cond}}}$. Using
the density of states in (\ref{EqSquareBand}) and integrating (\ref{EqSBTcond}%
) by parts in the weak-coupling limit, $D-\mu\gg k_{B}%
T_{\mbox{\scriptsize{cond}}}$, we get that \cite{Riseborough92}
\begin{equation}
k_{B}T_{\mbox{\scriptsize{cond}}}=\frac{2e^{\gamma}}{\pi}\left(  D-\mu\right)
\exp\left[  -D\frac{\mu-E_{f}}{V^{2}}\right]  \,, \label{EqSBTK}%
\end{equation}
where $\gamma$ is the Euler's constant and $N_{\sigma}=2$ in our model Hamiltonian.


\section{High temperature limit}

\label{High Temp}

\begin{figure}[tbh]
\centerline {\ \hspace{0.5cm}
\includegraphics[width=0.38\textwidth,
angle=90] {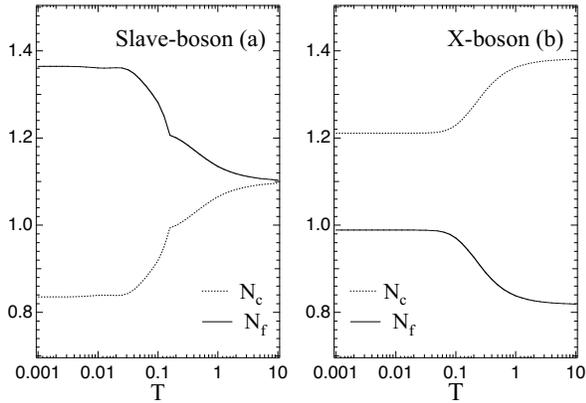} } \caption{Temperature
dependence of $N_{f}$ and $N_{c}$ for $N_{t}=2.2$ (a) in the
slave-boson approach and (b) in the X-boson approach. Other
parameters are the same as in Fig. \ref{FigSBV02zxT}.}
\label{FigNFandNCxT}
\end{figure}
In Section \ref{slave-boson} we have stated that for intermediate
values of $N_{t}$, but for $N_{t}<2$ (represented by $N_{t}=1.95$
in Fig. \ref{FigSBV02zxT}), $z$ is negative at a finite interval
only, becoming positive again at a sufficiently high temperature.
This follows in the MFSBT because $N_{f}=N_{c}$ in the high
temperature regime, and from $N_{t}=N_{f}+N_{c}<2$ we obtain
$N_{f}<1$, so that $z$ is positive.

First note that the occupations $N_{c}$ and $N_{f}$ are written as
\begin{equation}
N_{f,c}=N_{\sigma}\int d\epsilon\,\rho(\epsilon)\left[  A^{f,c}n_{F}%
(\omega_{+})+B^{f,c}n_{F}(\omega_{-})\right]  \,, \label{EqSBNc}%
\end{equation}
where
\begin{equation}
A^{f}=\frac{E_{\mathbf{k}}-\omega_{+}}{\omega_{-}-\omega_{+}}\;,\;B^{f}%
=\frac{\omega_{-}-E_{\mathbf{k}}}{\omega_{-}-\omega_{+}}\,, \label{EqSBAfBf}%
\end{equation}
\begin{equation}
A^{c}=\frac{\tilde{E_{f}}-\omega_{+}}{\omega_{-}-\omega_{+}}\;,\;B^{c}%
=\frac{\omega_{-}-\tilde{E_{f}}}{\omega_{-}-\omega_{+}}\,, \label{EqSBAcBc}%
\end{equation}
and $N_{\sigma}$ is the number of spin components per state $k$.

>From (\ref{EqSBPoles}) we can write $\omega_\pm=-\mu+\phi_\pm$,
with
\begin{equation}
\phi_\pm=\frac{1}{2}\left\{  \epsilon_{\mathbf{k}}+\left(  E_{f}%
+\lambda\right)  \pm\sqrt{\left[  \epsilon_{\mathbf{k}}-\left(  E_{f}%
+\lambda\right)  \right]  ^{2}+4\tilde{V}^{2}}\right\}  \,, \label{EqSBPoles2}%
\end{equation}
where $\phi_\pm$ are the true energies of the quasi-particles in
the Hamiltonian in (\ref{EqSBH}) before subtracting the quantity
$\mu\left\{
\sum_{\mathbf{j},\sigma}f_{\mathbf{j},\sigma}^{\dagger}f_{\mathbf{j},\sigma
}+\sum_{\mathbf{k},\sigma}c_{\mathbf{k},\sigma}^{\dagger}c_{\mathbf{k},\sigma
}\right\}  $. As the $\phi_\pm$ are bounded, we have $\lim_{\beta
\rightarrow0}\exp\left(  \beta\omega_\pm\right)  =\exp\left(
-\beta \mu\right)  $. Hence, in the high-temperatures limit,
(\ref{EqSBNc}) becomes
\begin{equation}
\lim_{\beta\rightarrow0}N_{c}=\lim_{\beta\rightarrow0}N_{f}=\frac{N_{\sigma}%
}{1+e^{-\beta\mu}}\,, \label{EqSBNtHT}%
\end{equation}
where we have made use of the relations
$A^{c}+B^{c}=A^{f}+B^{f}=1$, which can be verified by inspection
from (\ref{EqSBAfBf}) and (\ref{EqSBAcBc}). Further, from
(\ref{EqSBNtHT}) above, it follows immediately that $\mu$
satisfies
\begin{equation}
N_{t}=2N_{\sigma}\left(  1+e^{-\beta\mu}\right)  ^{-1}\label{EqSBMuHT}%
\end{equation}
in the high temperature limit.

Following the development above but with the GF corresponding to the MFXBT,
one can calculate $N_{f}$ and $N_{c}$ in the high-temperatures limit for the
X-boson approach. Indeed, from (\ref{EqXBPoles}), $\omega_\pm^{\sigma}%
=-\mu+\phi_\pm^{\sigma}$ and since $N_{f}\leq1$ in the MFXBT for every range
of temperatures and occupations, we find that $D_{\sigma}$ is bounded, so that
$\phi_\pm^{\sigma}$ is also bounded. Hence, $\lim_{\beta\rightarrow0}%
\exp\left(  \beta\omega_\pm^{\sigma}\right)  =\exp\left(  -\beta\mu\right)
$ and for $N_{\sigma}=2$ in the high-temperatures limit we have
\begin{equation}
\lim_{\beta\rightarrow0}N_{c}=\frac{2}{1+e^{-\beta\mu}}%
\;\;\mbox{ and }\;\;\lim_{\beta\rightarrow0}N_{f}=D_{\sigma}\,\frac
{2}{1+e^{-\beta\mu}}\,. \label{EqXBNtHT}%
\end{equation}
The presence of $D_{\sigma}$ in the GF given in (\ref{EqXBGf}) is reflected in
(\ref{EqXBNtHT}), and replacing $D_{\sigma}=1-(N_{f}/2)$ in this last equation
we find that
\begin{equation}
N_{f}=\frac{2}{2+e^{-\beta\mu}}\,, \label{EqXBNfHT}%
\end{equation}
which is identical to the result obtained in \cite{Franco03}.
Therefore, $N_{f}\not =N_{c}$ for the X-boson approach in the high
temperature limit. Indeed, Fig. \ref{FigNFandNCxT} shows the
temperature dependence of $N_{f}$ and $N_{c}$, for $N_{t}=2.2$. As
$T$ increases, $N_{f}\rightarrow1.1$ for the MFSBT, which is
unacceptable in the $U=\infty$ limit. This result does not occur
in the X-boson approach, as can be seen in Fig.
\ref{FigNFandNCxT}.b.

\end{document}